# Occupancy Estimation in Smart Buildings using Audio-Processing Techniques

Qian Huang[1], Zhenhao Ge[2], Chao Lu[3]

1) Assistant Prof., School of Architecture, Southern Illinois University Carbondale, Carbondale, IL, USA. Email: qhuang@siu.edu
2) Ph.D. Department of Electrical and Computer Engineering, Purdue University, West Lafayette, IN, USA. Email:zhenhao.ge@gmail.com
3) Assistant Prof., Department of Electrical and Computer Engineering, Southern Illinois University Carbondale, Carbondale, IL, USA. Email:chaolu@siu.edu

**Abstract:**

In the past few years, several case studies have illustrated that the use of occupancy information in buildings leads to energy-efficient and low-cost HVAC operation. The widely presented techniques for occupancy estimation include temperature, humidity, $CO_2$ concentration, image camera, motion sensor and passive infrared (PIR) sensor. So far little studies have been reported in literature to utilize audio and speech processing as indoor occupancy prediction technique. With rapid advances of audio and speech processing technologies, nowadays it is more feasible and attractive to integrate audio-based signal processing component into smart buildings. In this work, we propose to utilize audio processing techniques (*i.e.,* speaker recognition and background audio energy estimation) to estimate room occupancy (*i.e.,* the number of people inside a room). Theoretical analysis and simulation results demonstrate the accuracy and effectiveness of this proposed occupancy estimation technique. Based on the occupancy estimation, smart buildings will adjust the thermostat setups and HVAC operations, thus, achieving greater quality of service and drastic cost savings.

**Keywords:** smart building, occupancy estimation, audio processing, speaker recognition

## 1. INTRODUCTION

Nowadays, the research and innovation of smart buildings has been gaining increasingly attentions in building construction and management society. Among a variety of emerging technologies (*e.g.,* wireless sensor network, big data analytics, internet of things), demand driven HVAC (*i.e.,* heating, ventilation and air conditioning) operation is promising to significantly reduce the building operation cost (Agarwal, 2010) (Huang, 2010) (Yang, 2012), and ultimately to achieve sustainable smart buildings (Huang, 2011) (Huang, 2012) (Labeodan, 2015). Because HVAC facilities are the largest energy consumers in homes today, from some recently reported case studies, it has been found that the implementation of demand driven HVAC control results in 10% - 56% savings in building energy usage (Tachwali, 2007) (Erickson 2009) (Sun, 2011).

Room occupancy detection or estimation is of great importance to implement demand driven HVAC systems. Up to date, there are a couple of existing room occupancy detection techniques presented in literature (Sun, 2011) (Li, 2012) (Yang, 2012) (Wang, 2014) (Masoudifar, 2014). The basic idea of room occupancy detection or estimation is to install or activate various sensors, including $CO_2$ concentration sensor, RF ID sensor, motion sensor, light sensor, image camera, temperature sensor, acoustic sensor, humidity sensor, passive infrared sensor and so on. These sensors are supposed to collect occupancy information, which will be sent to building management system to control HVAC operation. Once the building management system is aware of the building occupancy information, it will adjust the setting parameters of HVAC facilities, such as air handler fan or air duct damper, to meet the demand of building users. In particular cases when no occupants stay in a room or thermal zone, heating and ventilation services could be stopped until the presence of occupants later. Therefore, instead of blindly running energy-consuming HVAC facilities, the acquired information of room occupancy enables to achieve a demand-driven HVAC operation and hence conserve energy.

As been stated and summarized in the papers (Yang, 2012) (Labeodan, 2015), these existing occupancy detection systems have some inherent drawbacks in terms of implementation cost, user privacy, detection accuracy and intrusiveness. Consequently, it is worthy to investigate other potential approaches to overcome these drawbacks of existing systems. With rapid advances of audio and speech processing technologies, nowadays it is more feasible and attractive to integrate audio-based signal processing components into smart buildings. Audio-based prototypes have been demonstrated in smart buildings to utilize speech recognition for building automation, such as voice-activated smart homes (Vacher, 2013) or voice controlled appliances for disabled or aging people (Suk, 2008).

To the best of our knowledge, the main research trends in smart homes in audio processing are related to speech recognition and human-smart home interaction. Yet, instead of further improving speech recognition algorithms, in this work, we propose to utilize audio processing techniques to estimate room occupancy. Theoretical analysis and simulation results for a case study is addressed in later sections. The proposed technique is cost-effective and non-intrusiveness, and also protects the user privacy very well. The feasibility and performance of room occupancy estimation are demonstrated to validate the proposed concept.

## 2. RELATED WORK

In the past few years, a variety of outstanding research achievements has been presented in the area of indoor occupancy detection. In 2008, researchers ever presented an occupancy detection solution using RF ID Sensors (Lee, 2008). Yet, it was revealed that the use of RF ID sensor results in coarse-grained detection performance (Li, 2012). Later, researchers proposed to use an image camera to calculate the number of room occupants (Erickson, 2009), or to install passive infrared sensors to detect occupancy of each room (Agarwal, 2010). The limitations of both approaches are high cost and user privacy concern (Li, 2012) (Labeodan, 2015). The measurement of $CO_2$ concentration in a room was proposed to determine the room occupancy (Sun, 2011) (Nassif, 2012). The main drawback of using $CO_2$ sensor is low accuracy, due to the output of $CO_2$ sensor heavily depends on the external conditions (Labeodan, 2015).

In terms of acoustic based research achievements, the researchers in 2013 firstly developed a sound detection system to run an occupancy detection algorithm (Uziel, 2013). This sound-based occupancy estimation algorithm investigated two different approaches: acoustic localization and machine learning algorithms. In 2014, an overview of three on-going projects which are supposed to use sound to reduce energy consumption in buildings is presented (Kelly, 2014). The idea of acquire, identify and monitor acoustic data in buildings is very attractive for energy efficient smart buildings. In 2014, researchers demonstrated an energy-efficient high-performance acoustic processing unit (Kattanek, 2014), which is an FPGA platform running audio processing algorithms.

Unfortunately, previous researches mainly focused on hardware description and implementation of acoustic based occupancy detection systems, so far there are little study available regarding specific audio-processing based occupancy techniques. To overcome this challenge, the focus of this study is to investigate the feasibility and performance of audio-processing techniques that can be utilized for indoor occupancy estimation in smart building environments. As we mentioned before, audio-based room occupancy processing is inherently cost-effective, since the required hardware is a microphone and a microcontroller, which is normally less than $10 US dollars. Audio-based occupancy detection does not require users to wear a RFID tag or light of sight, so it is non-intrusive and user-friendly. We believe it is highly demanding to create and develop smart audio-based occupancy estimation.

## 3. PROPOSED AUDIO-PROCESSING BASED OCCUPANCY ESTIMATION TECHNIQUE

### 3.1 Overview and Considerations

The estimation of room occupancy is based on audio recording through microphones. Two assumptions are made to restrict the scope of this investigation. First, the indoor collected sound is mainly human speech. No significant other audio sources are available, such as sounds from audio player and TV. Second, compared with sound level of the targeted rooms, the sound from neighboring rooms or outdoors is very weak, so it can be modeled and treated as Additive White Gaussian Noise (AWGN) with relative constant mean and small variance.

With the above two assumptions, the room occupancy estimation problem can be classified into two scenarios: meeting mode, in which each of the speakers speaks separately, and they can be distinguished from one another; and party mode, in which multiple persons are constantly speak at the same time.

In the meeting mode, the Principle Component Analysis/Linear Discriminative Analysis (PCA/LDA) based Gaussian Mixture Mode (GMM) speaker recognition system is used to recognize the voice of each speaker in the room (Ge, 2012), and then the number of speakers are summed up to determine the occupancy estimation. In the party mode, speaker recognition is not easy since speaker voices are mixed up and sounds like crowd noise. An important feature of speech, known as Short-Time Energy (STE), is used in this work to estimate the number of people. The details of audio processing techniques in the both modes will be addressed in following sections.

### 3.2 Occupation Estimation Based on Speaker Recognition

The PCA/LDA Gaussian Mixture Model (GMM) speaker recognition system uses one of the common types of speech features called Mel-Frequency Cepstral Coefficients (MFCCs), which converts overlapped 20 millisecond audio frames into high dimensional (*e.g.* 120 dimensions) feature vectors The dimensions of these features are firstly reduced using PCA, one of the common approaches for dimension reduction and feature enhancement (Ge, 2011). The features are then improved with discriminative training using LDA, one of the techniques for maximizing class separation (https://en.wikipedia.org/wiki/Linear_discriminant_analysis). As a consequence, the feature frames associated with each speaker are ready to construct a Gaussian Mixture Model, which represents the inherent characteristics of individual speaker. This stage is in learning phase. Later, when recognizing speakers from audio recordings, given the input PCA/LDA features of the testing speakers, the one which is most likely based on the learned GMM classifier will be selected as the hypothesis speaker.

$$\hat{S} = \arg \max_{s \in [1,S]} \Pr(\lambda_s | X) \tag{1}$$

where    $\Pr(\lambda_s | X)$: the posterior probability given the features of the testing speaker
        $\lambda_s$: the GMM speaker recognition model for speaker $s$

Generally, the more features (*i.e.,* the longer speech provided), the more accurate of recognition results are. Figure 1 demonstrates the speaker recognition performance with different number of Gaussian mixtures and different capacity of speaker pools. When an appropriate number of Gaussian mixtures are selected, the accuracy of speaker recognition is above 90% even when there are about 200 speakers in the room.

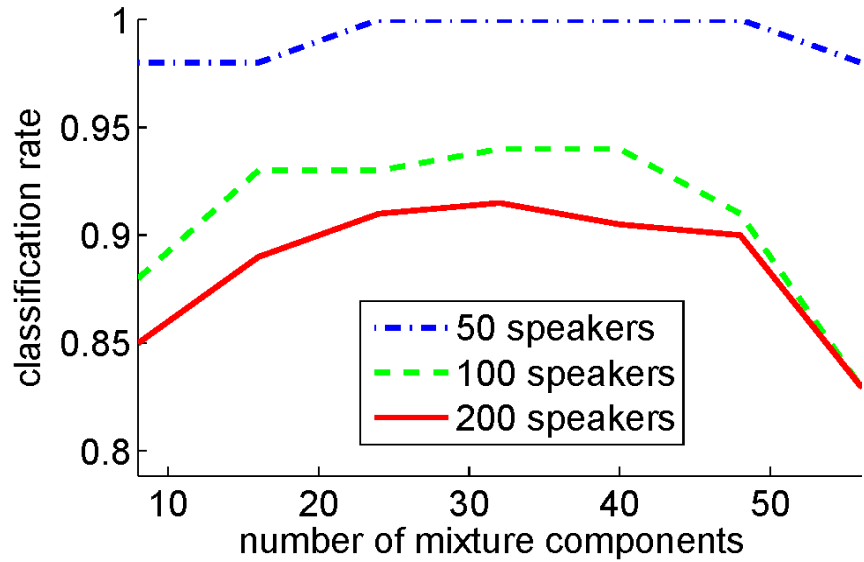

Figure 1. Speaker recognition accuracy vs. number of GMM mixtures for varying pool size

Compared with other speaking recognition systems available in the market, our processing technique based on PCA/LDA feature improvement is very robust and efficient. In our smart building application scenarios, the number of speaker is relatively small (*e.g.,* usually no more than 200 residents in a room). The details regarding extracting the MFCCs, implementing PCA/LDA feature improvement, and constructing GMM can be found (Ge, 2012).

**3.3 Occupation Estimation Based on Background STE**

In this section, we first generate random crowd speech with decay modelling; then, we estimate the STE level associated with different sizes of crowds; finally, we test the accuracy of occupation estimation based the STE levels.

(1) Crowd Speech Synthesis with Decay Modeling

There is no speech corpus available for crowd speech with Meta data, such as the number of speakers, relative positions of speaker with respect to the recording microphone. As a result, we artificially generate indoor background crowd speech by mixing single standard recording speech from different individual speakers using TIMIT corpus (Lopes, 2011).

In order to accurately model the indoor background speech and measure the STE of synthesized mixed speech, there are at least three factors, which may affect the background speech, should be well addressed. They include: (1) size and shape of the room changes, (2) position of the recording microphone, and (3) number of indoor people and their positions with respect to the microphone.

Rectangular or circular shaped rooms with central or corner single microphone positioning are modeled. Figure 2 presents how the speakers are positioned randomly in a room. The background speech recorded by the microphone is a summation of speeches from speakers with various distances. As shown in the sound distance law (www.sengpielaudio.com), the amplitude of speech is inversely proportional to the distance between the microphone to the sound source (*i.e.,* the mouth of speaker). Given the base amplitude $A_0$ recorded at distance $r_0$, the amplitude $A_i$ at distance $r_i$ should be

$$A_i = \frac{r_0}{r_i} A_0, where\ r_i > r_0 \tag{2}$$

Sound wave is assumed to evenly and circularly propagate towards every direction. In order to avoid large distortion of STE, due to some random positioned speakers stand too close to the location of microphone, the distance between speaker and microphone is required to be larger than $r_0$ (i.e., the small circle around microphone in Figure 2). This requirement makes sense in reality, since normally the microphone is well placed so that all speakers are kept away from the microphone with a certain distance. Furthermore, some advanced microphones have been integrated with overloud protection feature such as sound cutoff function.

With equation for sound decay in distance, the mixed speech recorded from the microphone can be expressed as

$$x = \sum_{i=1}^{N} A_i x_i = \sum_{i=1}^{N} \frac{r_0}{r_i} A_0 x_i = r_0 A_0 \sum_{i=1}^{N} \frac{x_i}{r_i} \tag{3}$$

where     $x_i$: the speech from $i_{th}$ speaker
             $N$: the number of speakers in the room

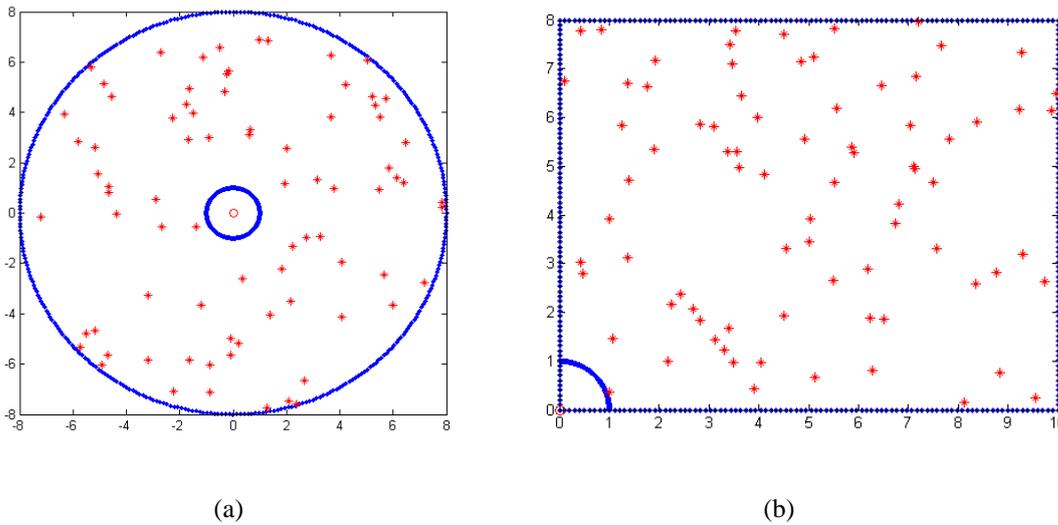

(a)                            (b)
Figure 2. (a) Random positions of speakers in a circular room with central microphone, (b) random positions of speakers in a rectangular room with corner microphone

In this study, the speaker density is limited to be no more than one speaker/square meter. It indicates that there are no more than 80 speakers in a rectangular room with a length of 10 meters and a width of 8 meters, as shown in Figure 2(b).

(2) STE Estimation

Short-Time Energy (STE) is a feature of signal energy among a short interval of time (*e.g.,* 50ms). Given *N* samples in short period, STE is calculated as

$$E(i) = \frac{1}{N} \sum_{n=1}^{N} w(n)\ |x_i(n)|^2$$

where     $x_i(n)$: the speech signal during $i_{th}$ frame segment
             $w(n)$: the windowing function to remove the artifacts along the edge of windowed signal (Giannakopoulos, 2014)

STE is usually measured frame by frame with overlaps. In this work, hamming window for window is used, the frame size is chosen as 50ms and step size is chosen as 25ms. For a given 5 second of speech, there are around 200 frames of STE measurement, as illustrated in Figure 3.

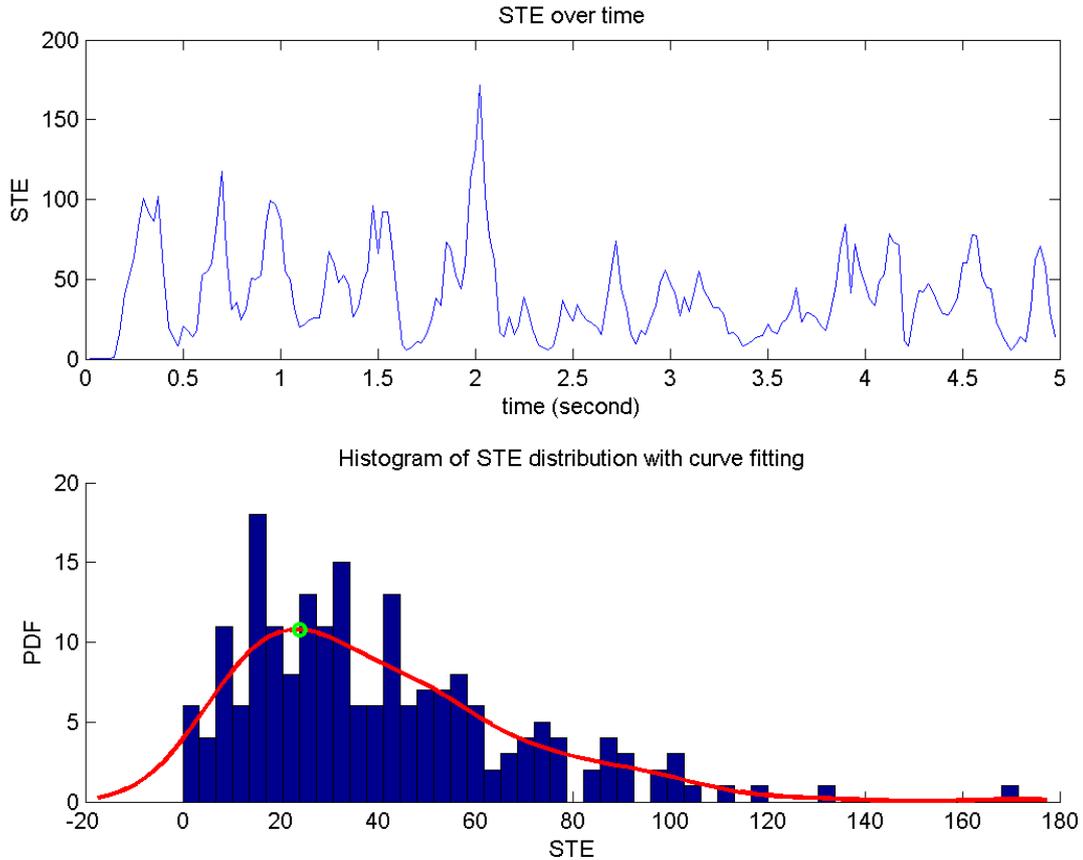

Figure 3. STE measurement and its statistical distribution for 10 random positioned speakers in a 10m×8m room

Based on this STE histogram, its Probability Density Function (PDF) can be obtained by curve fitting approach using nonparametric kernel-smoothing distribution in MATLAB. With the specific room shape and size, the STE measurement for a certain number of people in a room can be predicted based on Maximum a Posteriori (MAP) estimation, which is illustrated as the highest point (i.e., the green circle in Figure 3) of the PDF curve.

Given the room information and the number of occupants, the theoretically estimated STE $E_n$ is a constant value. Therefore, the room occupancy can be predicted by measuring $E_n$. However, the STE in reality follows a Gaussian distribution with a particular mean $\mu_n$ and variance σ. For example, Figure 3 shows a single observation of $E_n$ located at the green circle point, when the room occupancy is equal to 10.

Figure 4 illustrates the Gaussian distribution of STE with respect to various room occupancy. It is apparent that as the time of measurement is getting longer, the variance becomes smaller. Hence, the STE measurement for a particular room occupancy becomes more stable and relatively constant. Theoretical analysis also proves that the variance of PDF is inversely proportional to the room occupancy (*i.e.,* $\sigma_n^2 \sim 1/n$).

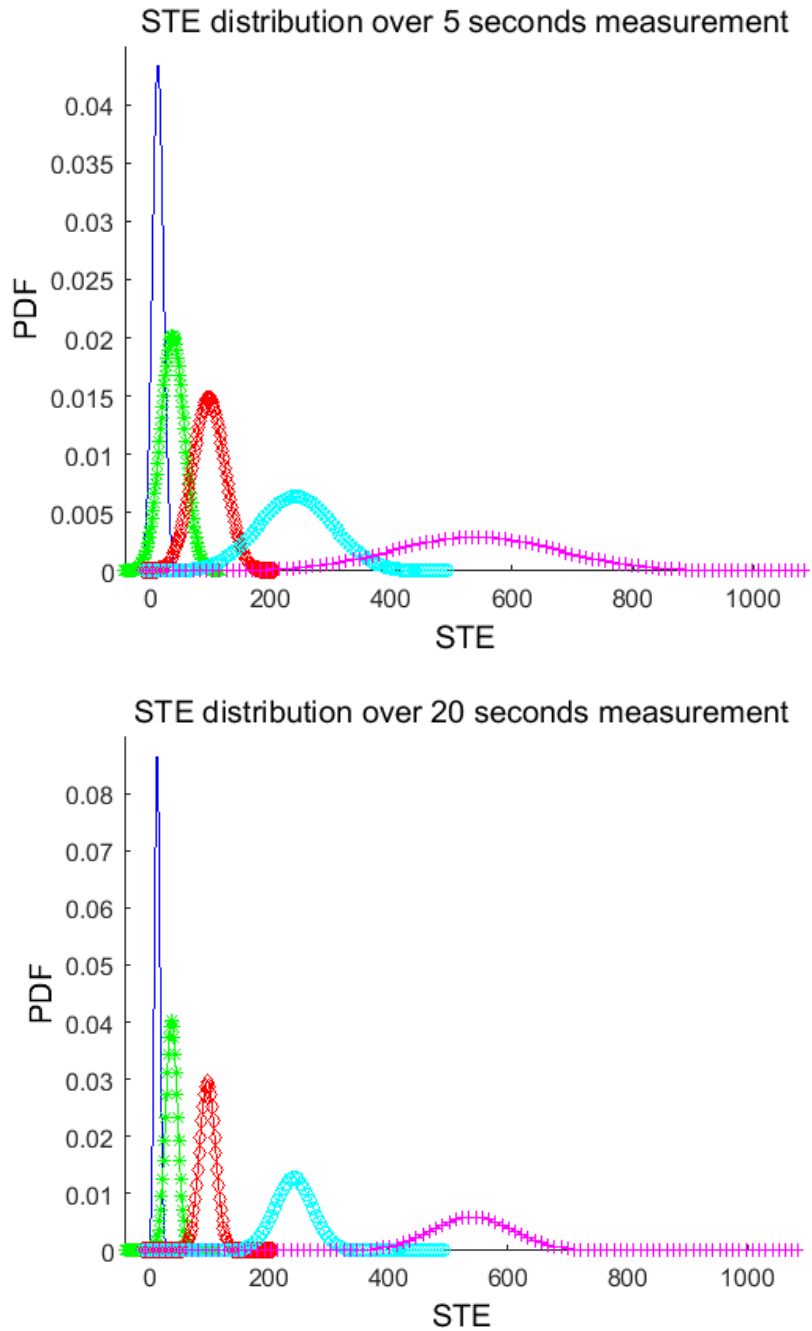

Figure 4. STE distribution with various crowd capacities (5-second and 20-second measurements)

(3) STE Based Room Occupancy Estimation Accuracy

After obtaining the estimated STE levels for various crowd sizes, the accuracy of estimation is tested over a series of different length of time periods. The crowd size with the maximum posterior probability is selected as the prediction value. The estimation results for a particular room setup (i.e., rectangular room with a length of 10 meters and a width of 8 meters) are shown in Table 1 and Figure 5.

We can see that the prediction accuracy can be enhanced with the increase of speech measurement time. The plot in Figure 5 shows the STE estimation is most challenging when the room occupancy size is medium, such as 10 or 20 speakers. This is reasonable since the probability distribution for these room occupancy values have more overlaps as shown in Figure 4. Even though there is some uncertainty due to overlaps of probability distribution curves, our proposed STE based estimation method realizes a decent accuracy prediction if the speech measurement time is longer than 20 seconds, as illustrated in Figure 5.

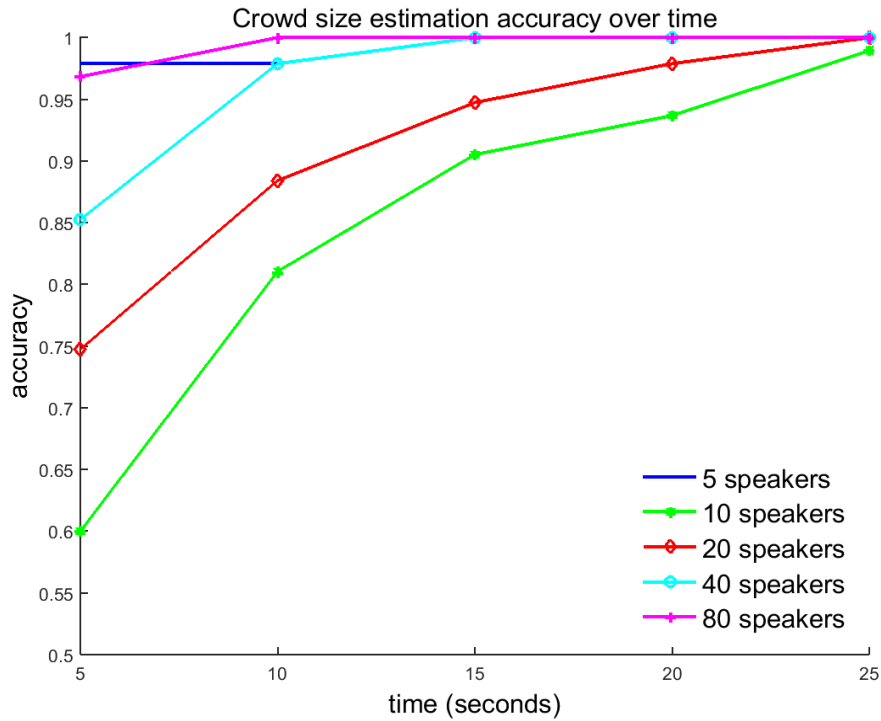

Figure 5. Crowd size estimation over time

Table 1. Estimation accuracy with various crowd sizes and speech measurement time

| Number of speakers\Time (seconds) | 5 | 10 | 15 | 20 | 25 |
|---|---|---|---|---|---|
| 5 | 0.98 | 0.98 | 1.00 | 1.00 | 1.00 |
| 10 | 0.60 | 0.81 | 0.91 | 0.94 | 0.99 |
| 20 | 0.75 | 0.88 | 0.95 | 0.98 | 1.00 |
| 40 | 0.85 | 0.98 | 1.00 | 1.00 | 1.00 |
| 80 | 0.97 | 1.00 | 1.00 | 1.00 | 1.00 |

## 4. CONCLUSIONS

Demand driven HVAC operation usually leads to significant savings of building energy consumption. With rapid advances of audio processing technologies, it is attractive to integrate audio-based occupancy estimation into smart buildings and enable demand driven HVAC operation. In this work, we propose two room occupancy estimation approaches, *i.e.,* for meeting mode and party mode, respectively. Theoretical models and design details are elaborated as well. Case studies and simulation results demonstrate very good estimation accuracy and hence validate its feasibility and benefits.